\documentclass[10pt,twocolumn,showpacs,longbibliography,amsmath,amssymb,prl,floatfix,superscriptaddress]{revtex4-2}

\usepackage{graphicx}
\usepackage{dcolumn}
\usepackage{bm}
\usepackage{color}
\usepackage{txfonts}
\usepackage{microtype}
\usepackage{epstopdf}

\usepackage{epstopdf}

\begin{document}

%\preprint{AIP/123-QED}

%\title{Controlling the polarization and vortex charge of $\gamma$ photons Driven by Bichromatic Circularly Polarized Laser Fields}
\title{Controlling the polarization and vortex charge of $\gamma$ photons via nonlinear Compton scattering}
\author{Jing-Jing Jiang}
\affiliation{Department of Physics, Shanghai Normal University, Shanghai 200234, China}
\author{Kai-Hong Zhuang}
\affiliation{Department of Physics, Shanghai Normal University, Shanghai 200234, China}
\author{Jia-Ding Chen}
\affiliation{Department of Physics, Shanghai Normal University, Shanghai 200234, China}
\author{Jian-Xing Li}
\email{jianxing@xjtu.edu.cn}
\affiliation{Ministry of Education Key Laboratory for Nonequilibrium Synthesis and Modulation of Condensed Matter,  State key laboratory of electrical insulation and power equipment,  Shaanxi Province Key Laboratory
of Quantum Information and Quantum Optoelectronic Devices,  School of Physics, Xi’an Jiaotong University, Xi’an 710049, China}
\affiliation{Department of Nuclear Physics, China Institute of Atomic Energy, P.O. Box 275(7), Beijing 102413, China}
\author{Yue-Yue Chen}
\email{yue-yue.chen@shnu.edu.cn}
\affiliation{Department of Physics, Shanghai Normal University, Shanghai 200234, China}
\date {\today}

\begin{abstract}
High-energy vortex $\gamma$ photons have significant applications in many fields, however, their generation and angular momentum manipulation are still great challenges. Here, we first investigated the generation of vortex $\gamma$ photons with controllable spin and orbital angular momenta via nonlinear Compton scattering of two-color counter-rotating circularly polarized (CP) laser fields. 
%The radiation probability of twisted photons [O. V. Bogdanov, \textit{et al.} PRD 99, 116016 (2019)] are employed to predict the angular momentum  of photons from charged particle trajectories. 
The radiation probabilities of vortex photons are calculated using the semiclassical approach that resolves angular momenta of emitted photons. % and allows for the investigation of angular momentum transfer in arbitrary laser fields. 
We find that % the electron absorbs a combination of photons with opposing spin angular momenta. 
the angular momenta transferred to emitted photons are determined by the dominating photon absorption channel, leading to a structured spectrum with alternations in {\color{black}{helicity}} and twist directions. 
By tuning the relative intensity ratio of the two-color CP laser fields, %the preferred photon absorption channel can be enhanced, providing 
the polarization and vortex charge of the emitted $\gamma$ photons can be controlled, enabling
%the photons with either the negative or positive spin (orbital) angular momentum can be preferentially selected. 
 the generation of circularly polarized vortex $\gamma$ photons with a user-defined polarization and topological charge, which may have a plenty of applications in nuclear physics, astrophysics, particle physics, etc. % Meanwhile, for a certain opening angle, we demonstrate a spectral spin (orbital) oscillation governed by the conservation laws of energy and angular momentum. The spin angular momentum alternates between left- and right-handed circular polarization as the photon energy increases, while orbital angular momentum alternates between positive and negative topological charge.  This unique structure enables the generation of spatially isolated vortex beams of opposite spin and orbital angular momentum, which may have applications in nuclear spin probing and control.
\end{abstract}

\maketitle

%\section{Introduction}
{\color{black}{The vortex photons are the states of the electromagnetic field that possess a definite energy, unique projections of the momentum and of the total angular momentum onto the detector axis, and a well defined helicity \cite{ivanov2022promises,ivanov2011colliding,jentschura2011compton,jentschura2011generation}.}} High-energy vortex photons are opening new avenues in high-energy and nuclear physics. 
%High-energy vortex photons, carrying orbital angular momentum (OAM), are opening new avenues in high-energy and nuclear physics.
These photons are predicted to induce novel effects in photoionization \cite{picon2010photoionization}, violate dipolar selection rules \cite{picon2010photoionization}, and uncover quadrupolar transitions in x-ray magnetic circular dichroism \cite{van2007prediction}. %In resonant inelastic x-ray scattering, they provide insights into molecular structures via vibrational resonances \cite{rury2013examining}. 
In nuclear physics, $\gamma$ vortices enable new methods to study hadron spin-parity properties \cite{ivanov2020doing} and enhance nuclear isomer depletion \cite{wu2022dynamical,lu2023manipulation}. The interactions of vortex photons with ion beams offer promising experimental possibilities, positioning vortex photons as powerful tools for advancing nuclear and particle physics \cite{ivanov2020doing,ivanov2022promises,durante2019all,krasny2015gamma,budker2020atomic,budker2022expanding}. Additionally, $\gamma$ vortex photons are key to probing quantum electrodynamics (QED) effects, influencing angular distributions and polarization in Compton scattering \cite{stock2015compton,maruyama2019compton,sherwin2017theoretical,sherwin2017compton}, as well as unveiling novel interactions in photon-photon scattering and positron production \cite{aboushelbaya2019orbital,lei2021generation,bu2021twisted}. $\gamma$ vortices with tunable spin angular momentum (SAM) and orbital angular momentum (OAM) could greatly expand their applications, offering unprecedented detection accuracy and control over interactions. 

To date, several schemes for generating bright GeV $\gamma$-photon beams with controllable angular momentum have been proposed, employing nonlinear Compton scattering (NCS) with intense CP Gaussian or Laguerre-Gaussian laser pulses \cite{ju2019generation,zhu2018generation}. Unfortunately, these schemes use a semiclassical approach, modelling electron motion classically while treating emission quantum mechanically as plane-wave. This approach enables analysis of the collective OAM of $\gamma$-ray beams via $\vec{L}=\sum_i \vec{r}_i\times \vec{p}_i$, with $\vec{r}_i$ and $\vec{p}_i$ being the position and momentum of the individual photons, but it does not capture intrinsic OAM. A more rigorous quantum treatment defines the emission as Bessel states, characterized by definite energy, longitudinal momentum projection, total angular momentum projection, and helicity \cite{jentschura2011generation,molina2007twisted,jentschura2011generation,ivanov2011colliding,jauregui2005quantum,bialynicki2006beams}.  The QED cross section for Compton scattering has been calculated with the eigenfunction for single photons carrying OAM \cite{jentschura2011generation,jentschura2011compton,stock2015compton}. Recent developments include extending this framework to the nonlinear regime using laser-dressed vortex states \cite{bu2023twisting,ababekri2024vortex}. Additionally, the semiclassical method has been updated to treat $\gamma$-photons as vortex states \cite{bogdanov2019semiclassical,bogdanov2018probability}. This modification allows for the calculation of the radiation probability of vortex $\gamma$ photons by integrating electron trajectories, offering an approach for studying vortex radiation in arbitrary laser field configuration. Despite these advances in formalism, an effective scheme for controlling the angular momentum of $\gamma$-vortex photons has yet to be developed.

Two-color counter-rotating circularly polarized (CP) laser fields have been extensively employed to control the polarization and vortex charge of attosecond high-harmonic beams through simultaneous spin–orbit momentum conservation \cite{kong2017controlling,dorney2019controlling,
gariepy2014creating,gauthier2017tunable}.  %This approach enables manipulation of the polarization of attosecond pulses and the generation of circularly polarized vortices with tailored orbital angular momentum. 
Utilizing the two-frequency fields in energetic upconversion of photons could provide better control of electrodynamics and emission spectrum, opening possibility for achieving $\gamma$ vortex with tunable angular momentum. However,  previous efforts to generate $\gamma$ vortices have been restricted to NCS with single-{\color{black}{helicity}} fields, e.g. single-wavelength CP fields  \cite{ababekri2024vortex} or two-wavelength co-rotating CP laser fields \cite{taira2018gamma}. In such fields, the angular momentum transfer to the system has a definitive sign, predetermined by the helicity of the laser,  which restricts control over the {\color{black}{helicity}} and twist directions of the emitted $\gamma$ photons.  {\color{black}{Here, helicity direction refers to the alignment of the photon’s spin with its propagation, while twist direction refers to the rotational direction of the photon’s wavefront.}} Thus, the manipulation of SAM and OAM of $\gamma$ photons remains a great challenge.

\begin{figure}
    \includegraphics[width=0.5\textwidth]{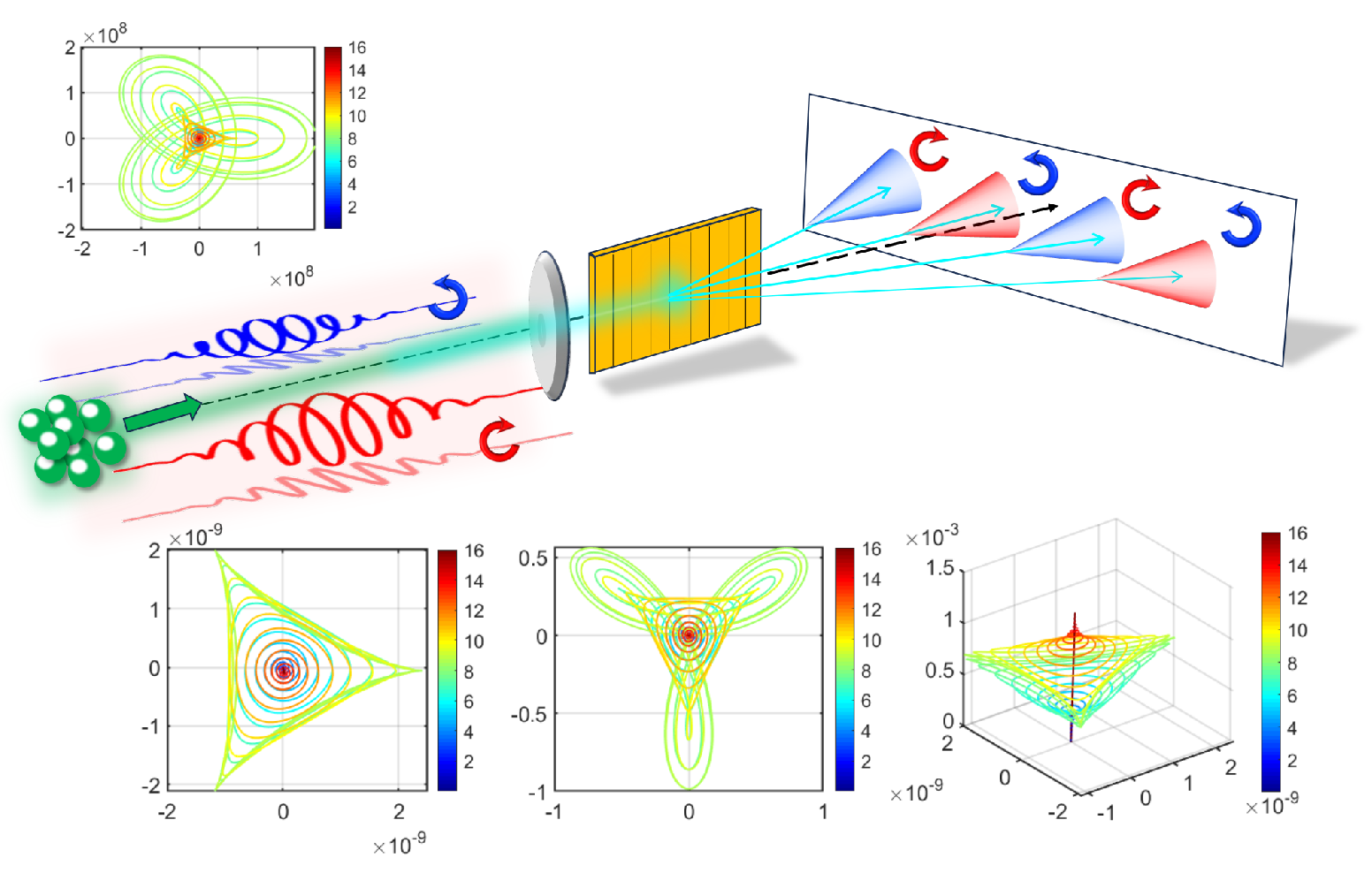}
       \begin{picture}(300,15)  
       \put(5,80){{\fontsize{6pt}{\baselineskip}\selectfont$\mathbf{e^-}$}} 
        \put(73,133){{\fontsize{6pt}{\baselineskip}\selectfont{\color{blue}$\mathbf{2\omega_0}$, \textbf{LCP}}}} 
        \put(100,90){{\fontsize{6pt}{\baselineskip}\selectfont{\color{black}$\mathbf{\omega_0}$, \textbf{RCP}}}}     
         \put(113,136){{\fontsize{6pt}{\baselineskip}\selectfont \rotatebox{15}{\textbf{Grating}}}}
             \put(175,163){{\fontsize{6pt}{\baselineskip}\selectfont \rotatebox{-11}{\textbf{vortex $\gamma$ photons}}}}
     \put(150,165){\fontsize{7pt}{\baselineskip}\selectfont (a)}
    \put(16,165){\fontsize{4.5pt}{\baselineskip}\selectfont (b)}
    \put(70,33){\fontsize{5.0pt}{\baselineskip}\selectfont (c)}
      \put(145,33){\fontsize{5.0pt}{\baselineskip}\selectfont (d)}
       \put(220,41){\fontsize{5pt}{\baselineskip}\selectfont (e)}
        \put(60,173){{\fontsize{4.5pt}{\baselineskip}\selectfont${t/T}$}}
        \put(35,125){{\fontsize{4.5pt}{\baselineskip}\selectfont${E_x}$}}
        \put(3,150){\fontsize{4.5pt}{\baselineskip}\selectfont\rotatebox{90}{\fontsize{4.5pt}{\baselineskip}\selectfont$E_y$}}
        \put(45,19){{\fontsize{6.0pt}{\baselineskip}\selectfont$x$ (cm)}}
        \put(20,46){\fontsize{6.0pt}{\baselineskip}\selectfont\rotatebox{90}{\fontsize{6.0pt}{\baselineskip}\selectfont$y$ (cm)}}
        \put(83,80){{\fontsize{5pt}{\baselineskip}\selectfont${t/T}$}}
         \put(126,19){{\fontsize{5.0pt}{\baselineskip}\selectfont$P_x$}}
        \put(93,53){\fontsize{5.0pt}{\baselineskip}\selectfont\rotatebox{90}{\fontsize{5.0pt}{\baselineskip}\selectfont$P_y$}}
         \put(155,80){{\fontsize{5pt}{\baselineskip}\selectfont${t/T}$}}
        \put(213,23){{\fontsize{6.0pt}{\baselineskip}\selectfont$x$ (cm)}}
        \put(178,24){{\fontsize{6.0pt}{\baselineskip}\selectfont$y$ (cm)}}
        \put(166,49){\fontsize{6.0pt}{\baselineskip}\selectfont\rotatebox{90}{\fontsize{6.0pt}{\baselineskip}\selectfont$z$ (cm)}}
        \put(235,84){{\fontsize{5pt}{\baselineskip}\selectfont${t/T}$}}
      \end{picture}
      \vskip -0.6cm
    \caption{(a) Scheme for generating vortex photons with two-color counter-rotating CP laser fields. For a specific opening angle, the SAM (OAM) reverses sign as the harmonic order increases. {\color{black}{The harmonics can be spatially separated using a crystal diffraction gratings \cite{smither1995review,barriere2009experimental}, or alternatively a Compton-based $\gamma$ ray spectroscopy \cite{eberl2003nuclear}.}} The time evolution of laser electric field (b), electron trajectory  (c), and momentum (d), projected in the polarization plane ($x-y$). (e) The three-dimensional time evolution of electron trajectory. The color indicates the evolution time in unit of laser period. }
    \label{Fig. scheme}
\end{figure}

In this Letter, we investigate the simultaneously manipulation of SAM and OAM of emitted $\gamma$ photon in NCS at moderate intensities \cite{yoon2021realization}; see Fig. \ref{Fig. scheme} (a).  Employing the semi-classical operator method recently refined %by O. V. Bogdanov \textit{et al.} 
to resolve photon angular momentum \cite{bogdanov2019semiclassical}, we obtained the radiation probability of vortex $\gamma$ photons. % in a two-color counter-rotating CP laser field.
In our two-color scheme, 
the electron absorbs a combination of photons with opposing SAM, resulting in a total angular momentum with a tunable sign dictated by the dominant photon absorption channel.
We find that left-circular (LCP) or right-circular polarization (RCP) harmonics, which possess either a  positive or negative topological charge {\color{black}{(OAM value)}} are generated throughout the spectrum; see Figs. \ref{Fig. scheme} (a) and {\color{black}{\ref{Fig. m}. As a key result, vortex $\gamma$ photons with controllable SAM and OAM are generated; see Fig. \ref{Fig. scaling}. In contrast to previous works \cite{ju2019generation,zhu2018generation,
ababekri2024vortex,bogdanov2019semiclassical}, we demonstrate that, }}
by simply manipulating the relative intensity ratio of the two-color fields, one can select the preferred photon absorption channels, 
enabling control over the {\color{black}{helicity}} and twist directions of the $\gamma$ radiation. Our study bridged the photonic gap between angular momentum control in the extreme ultraviolet and $\gamma$-ray regimes,  enabling the manipulation of both the SAM and OAM of $\gamma$ photons from negative to positive values.
The generated $\gamma$-vortex photons with tailored angular momentum hold promising applications in nuclear physics, laboratory astrophysics, and particle physics, {\color{black}{offering promising implications for future NCS experimental campaigns involving OAM $\gamma$-rays \cite{chen2024platform}.}}

%\section{Scheme and Simulation methods}
 
We consider a ultra-relativistic electron with energy $\varepsilon_{0} = 1$ GeV colliding with an intense two-color CP laser pulse; see Fig. \ref{Fig. scheme}. 
The laser pulse propagates $-z$ direction while the electron propagate along $z$. 
The field consists of two co-propagating laser pulses of $\lambda_1=0.8\mu$m and $\lambda_2=0.4\mu$m wavelengths, $\tau_{p1}=10T_1$ and  $\tau_{p2}=20T_1$ pulse durations, right-circular and left-circular polarization, respectively. %The beam waist size is $5\lambda_1$ for both laser pulses. 
The peak amplitudes of each field, $E_{01}$ and $E_{02}$, fulfil the ratio $E_{01}/E_{02}=1/2$, i.e.,  $a_{01}/a_{02}=1$ for the field parameters. Here, $a_{0}=eE_{0}/m\omega_{0}$, where $\omega_{0}$ represents the laser frequency. We choose a laser intensity of $I_{01}\approx 5\times10^{17}\text{W/cm}^2$ ($a_{01}=0.5$), which is available  with current laser technology \cite{dorney2019controlling}. The Gaussian pulse shape is given by $g(\varphi)=\exp{\left(\frac{\varphi^2}{2\varphi_m^2}\right)}$ with $\varphi_m=\frac{\omega_{0}\tau_{p}}{2\sqrt{2\ln2}}$ and $\varphi$ being the laser phase.  The cloverleaf-like electric field in polarization plane are shown in Fig. \ref{Fig. scheme} (b). {\color{black}{ The considered NCS process can be expressed with
\begin{equation}
e + n_{R} \gamma_{R} + n_{L} \gamma_{L} \rightarrow e' + \gamma,
\end{equation}
where \( \gamma_R \) and \( \gamma_L \) represent the RCP and LCP laser photons, respectively; \( n_{R,L} \) are the corresponding photon numbers involved in the interaction, and \( \gamma \) denotes the emitted high-energy vortex photon.}}
The strong-field quantum parameter \( \chi_e = \left| F_{\mu \nu} p^{\nu} \right| / m F_{\text{cr}} \) governs the strength of radiation reaction and the OAM imparted to the scattered electron \cite{bu2023twisting}. Under the conditions considered, \( \chi_e \ll 1 \) {\color{black}{(\(a_0 \lesssim 10^4/\gamma_0 \) with $\gamma_0=\varepsilon_0/m$, see more details in Appendix A)}}, implying that the electron primarily serves as an intermediary, channelling angular momentum from the laser to the photons. This allows for an effective treatment of the electron as a plane wave, with the photon adopting a vortex state.

  \begin{figure}[b]
    \includegraphics[width=0.5\textwidth]{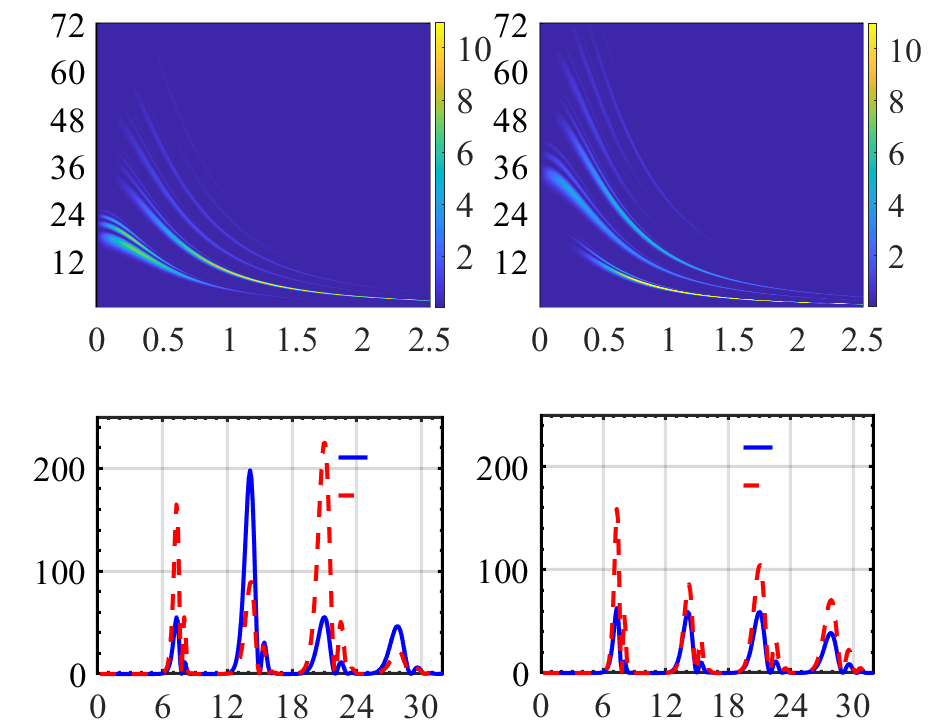}
    \begin{picture}(300,15)       
     \put(31,198){\color{white}(a)}
     \put(61,108){\fontsize{9pt}{\baselineskip}\selectfont$\theta_\perp$ (mrad)}
     \put(1,149){\fontsize{9pt}{\baselineskip}\selectfont\rotatebox{90}{$\omega$ (MeV)}} 
     \put(153,198){\color{white}(b)}
     \put(175,108){\fontsize{8.5pt}{\baselineskip}\selectfont$\theta_\perp$ (mrad)}
     \put(31,91){(c)} 
     \put(61,6){\fontsize{9pt}{\baselineskip}\selectfont$\omega$ (MeV)} 
     \put(1,43){\fontsize{8.5pt}{\baselineskip}\selectfont\rotatebox{90}{\fontsize{8.5pt}{\baselineskip}\selectfont$\omega$dP/$d\omega$}}  
      \put(104,88){\fontsize{7.5pt}{\baselineskip}\selectfont{s=+1}} 
      \put(104,77){\fontsize{7.5pt}{\baselineskip}\selectfont{s=-1}} 
     \put(153,91){(d)}
     \put(178,6){\fontsize{9pt}{\baselineskip}\selectfont$\omega$ (MeV)}
      \put(215,91){\fontsize{7.5pt}{\baselineskip}\selectfont{s=+1}} 
      \put(215,80){\fontsize{7.5pt}{\baselineskip}\selectfont{s=-1}}
     
    \end{picture}
    \vskip -0.2cm
    \caption{The angle-resolved energy spectra $\omega \text{d}^2P/\text{d}\omega \text{d}\theta_\perp$  (rad$^{-1}$) for photon helicities (a) $s=+1$ and (b) $s=-1$. (c) The energy spectra $\omega \text{d}P/\text{d}\omega$  for the vortex $\gamma$ photons with $\theta_\perp=0.72$ mrad for $s=+1$ (blue solid line) and $s=-1$ (red dashed line). (d) Same as (c) but for a co-rotating two-color CP fields. The laser intensities are  $a_{01}=a_{02}=0.5$, and electron energy $\varepsilon_0=1$ GeV.}
    \label{Fig. agl}
\end{figure}

\begin{figure}[t]
    \includegraphics[width=0.5\textwidth]{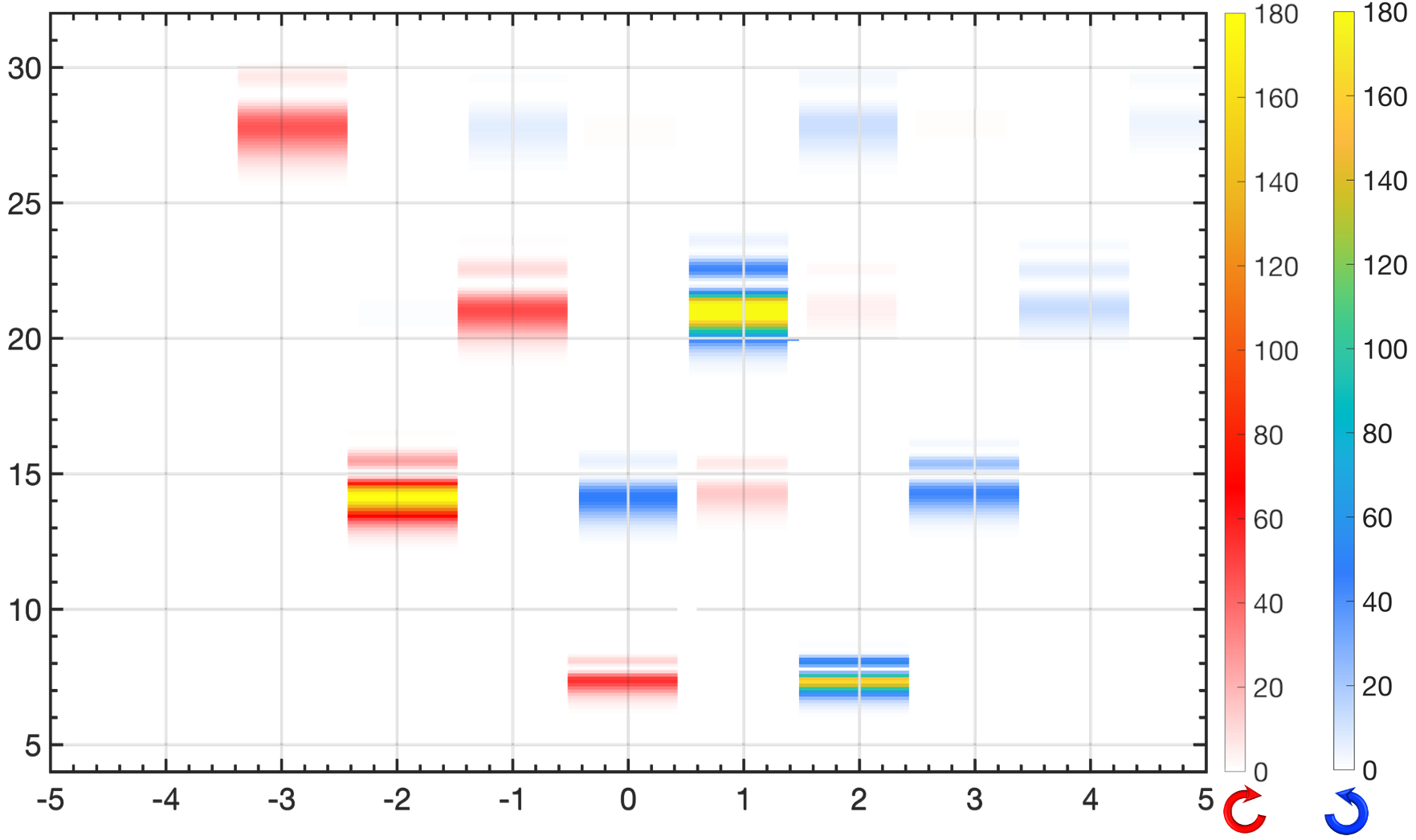}
    \begin{picture}(300,20)
    \put(120,10){\fontsize{9pt}{\baselineskip}\selectfont{${l}$}}
     \put(-10,85){\fontsize{9pt}{\baselineskip}\selectfont\rotatebox{90}{$\omega$ (MeV)}} 
    \end{picture}
    \vskip -0.5cm
    \caption{The OAM-resolved energy spectra $\omega \text{d}P/\text{d}\omega$   for photon helicities $s=+1$ (hot) and $s=-1$ (parula). The opening angle is $\theta_\perp=0.72$ mrad. The parameters are the same as in  Fig. \ref{Fig. agl}. }
    \label{Fig. m}
\end{figure}

%The radiation probability of twisted photons is numerically calculated by integrating the trajectory of the electron with the semiclassical formula provided in Refs. \cite{bogdanov2019semiclassical,bogdanov2018probability}.
In the ultrarelativistic limit, the one-photon radiation probability can be calculated by means of a formula resembling the classical formula for the intensity of radiation \cite{baier1998electromagnetic}. Instead of solving the Dirac equation in the specified external fields, it suffices to find the solutions to the Lorentz equations within these fields \cite{wistisen2019numerical,lv2022high}. %The semiclassical method treats the electron motion classically while considering photon emission quantum mechanically. 
The radiation probability of vortex photons can be derived using the Baier-Katkov method, by integrating the plane wave formula over the azimuthal angles of the photon momenta, incorporating the appropriate weights \cite{bogdanov2019semiclassical}. 
When quantum recoil is negligible, the radiation probability  of a vortex photon for ultrarelativistic particles moving in a general electromagnetic field reads \cite{bogdanov2019semiclassical,bogdanov2018probability}
\begin{align}\label{probability}
&dP\left(s,m,k_{3},k_{\perp}\right)=\alpha\left|I_{3}+\frac{1}{2}\left(I_{+}+I_{-}\right)\right|^{2}n_{\perp}^{3}\frac{dk_{3}dk_{\perp}}{4\pi},
\end{align}
where 
\begin{align*}
&I_{3}=\int dt\dot{x}_{3}e^{-i\omega t\left(1-n_{3}\upsilon_{3}\right)}j_{m}\left(k_{\perp}x_{+},k_{\perp}x_{-}\right)\\
&I_{\pm}=\frac{in_{\perp}}{s\mp n_{3}}\int dt\dot{x}_{\pm}e^{-i\omega t\left(1-n_{3}\upsilon_{3}\right)}j_{m\mp1}\left(k_{\perp}x_{+},k_{\perp}x_{-}\right).
\end{align*}
%Here $\omega$ is the twisted photon energy (units with $\hbar=c=1$ are employed throughout the paper). 
Here the quantum number $s$ defines the helicity of the vortex photon, $m$ the projection of the total angular momentum onto $\mathbf{e}_3$. {\color{black}{The unit vector $\mathbf{e}_3$ is the quantization axis for angular momentum, pointing from the radiation source to the detector (see more details in Appendix B).}} The function $j_m$ is related to the Bessel function of the first kind.  In our simulation, we numerically solve the Lorentz equations to obtain the trajectory of the electron, and then substitute the trajectory into Eq. (\ref{probability}) to calculate the probability of vortex photon emission. The accuracy of our code has been benchmarked with the QED calculations in Ref. \cite{ababekri2024vortex} (see more details in Appendix C).
%Here the quantum number $s$ defines the helicity of the vortex photon, $m$ the projection of the total angular momentum onto $\mathbf{e}_3$, $k_3$ and $k_\perp$ the projection of the photon momentum $k$ onto and orthogonal to $\mathbf{e}_3$, respectively.  We define $n_3=k_3/k$ and $n_\perp=k_\perp/k$. The unit vector $\mathbf{e}_3$  points from the radiation source to the detector. In this work, we choose  $\mathbf{e}_3=(0,0,1)$ along the collision axis $z$, with the orthonormal vectors $\mathbf{e}_1=(1,0,0)$ and $\mathbf{e}_2=(0,1,0)$, forming a right-handed coordinate system. The coordinates are represented as $\bm{x}_\pm(t)=(\mathbf{e}_\pm,\bm{x})$  and the velocities as $\bm{\upsilon}_\pm(t)=(\mathbf{e}_\pm,\bm{\upsilon})$, where $\mathbf{e}_\pm=\mathbf{e}_1\pm i\mathbf{e}_2$. The function $j_m$ is related to the Bessel functions of the first kind $J_m$ through the relation $j_{m}\left(p,q\right)=\frac{p^{m/2}}{q^{m/2}}J_{m}\left(\sqrt{p},\sqrt{q}\right)$. In our simulation, we numerically solve the Lorentz equations to obtain the trajectory of the electron, and then substitute the trajectory into Eq. (\ref{probability}) to calculate the probability of vortex photon emission. The accuracy of our code has been benchmarked with the QED calculations in Ref. \cite{ababekri2024vortex} (see more details in Appendix).

%\section{Simulation results}

The simulation results of the angle-resolved energy spectra are shown in Figs. \ref{Fig. agl} (a) and (b). Both RCP and LCP spectra exhibit multiple curved strips, which represent different harmonic orders. For each strip, the opening angle of the vortex photon $\theta_\perp=k_\perp/k$, increases with the decreases of emitted photon energy \cite{jackson2021classical,blackburn2020radiation}.
%, roughly follows $\theta_\perp\sim \delta^{-1/3}$ for $\delta\ll1$ \cite{jackson2021classical}, and $\theta_\perp\sim \sqrt{(1-\delta)/\delta}$ for $\delta\sim1$ \cite{blackburn2020radiation}, where $\delta=\omega/\varepsilon_0$. 
%Within each harmonic, sidebands appear due to ponderomotive broadening resulting from the finite pulse shape \cite{seipt2016analytical}. 
Furthermore, compared with radiation in an RCP field (see Fig. \ref{Fig. benchmark} in Appendix), the introduction of second harmonic LCP laser photons causes $s=1$ component to extend into the large-angle and small-energy region [\ref{Fig. agl} (a)], while $s=-1$ photons shift into the small-angle and large-energy region [\ref{Fig. agl} (b)], leading to a more pronounced overlap of $s=1$ and $s=-1$ photons in the angle-resolved energy spectra.
%Since the vortex $\gamma$ photon to be absorbed by the nuclear or atomic system is in a coherent state with definite opening angle $\theta_\perp$, we present the energy spectra for a fixed angle in Fig. \ref{Fig. agl} (c). The spectrum at $\theta_\perp=0.72$ mrad is dominated by the first three harmonics. For Subpeaks appear alongside the main peaks, corresponding to the sidebands shown in Fig. \ref{Fig. agl} (a) and (b). More interesting,  
More interesting, at $\theta_\perp=0.72$ mrad, the RCP and LCP alternate dominance throughout the spectrum: the LCP leads the RCP in the first and third harmonics, while the RCP takes the lead in the second and fourth harmonics [Fig. \ref{Fig. agl} (c)]. %This shift of leading polarization enables selective enhancement of certain polarizations and allows for complex manipulation of nuclear or atomic systems. 
In contrast, the polarization pattern in co-rotating two-color CP fields shows RCP dominating over LCP for all harmonics [Fig. \ref{Fig. agl} (d)]. 
This occurs because the photon emissions in an RCP laser field with helicity $s=-1$ have a larger opening angle $\theta_\perp$ compared to photons with helicity $s=1$, leading to the  the dominance of LCP radiation at $\theta_\perp=0.72$ mrad throughout the spectrum. In this case, the electron absorbs $m=n_1+n_2$ angular momentum and generates LCP $\gamma$ photons with positive topological charge $l=n_1+n_2-1$. 
  
\begin{figure}[b]
    \includegraphics[width=0.48\textwidth]{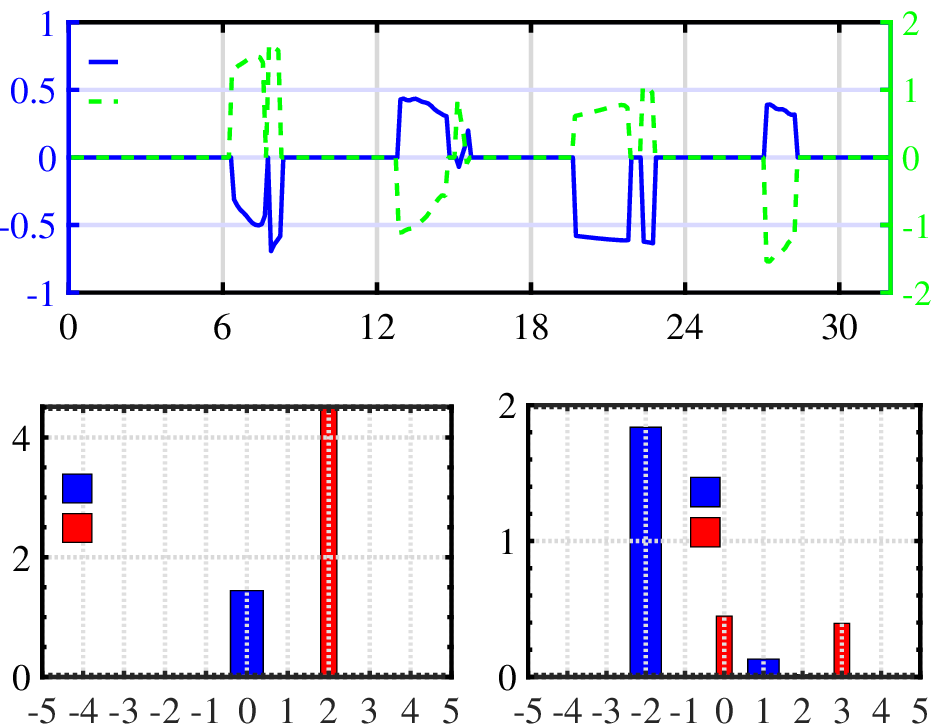}
    \begin{picture}(300,15)       
     \put(218,191){(a)}
     \put(34,188){\fontsize{8.5pt}{\baselineskip}\selectfont{$\overline{\xi}_2$}}
     \put(33,175){\fontsize{8.5pt}{\baselineskip}\selectfont{$\overline{l}$}}
     \put(103,108){\fontsize{8.5pt}{\baselineskip}\selectfont{$\omega$ (MeV)}}
     \put(104,90){(b)}
     \put(66,7){\fontsize{8.5pt}{\baselineskip}\selectfont{${l}$}}
      \put(30,76){\fontsize{8.5pt}{\baselineskip}\selectfont{s=+1}}
       \put(30,64){\fontsize{8.5pt}{\baselineskip}\selectfont{s=-1}}
      \put(226,91){(c)}
     \put(172,7){\fontsize{8.5pt}{\baselineskip}\selectfont{${l}$}}
     \put(193,75){\fontsize{8.5pt}{\baselineskip}\selectfont{s=+1}}
       \put(193,63){\fontsize{8.5pt}{\baselineskip}\selectfont{s=-1}}
      \put(-6,49){\fontsize{8.5pt}{\baselineskip}\selectfont\rotatebox{90}{\fontsize{8.5pt}{\baselineskip}\selectfont {dP/$d\omega$}}}   
     \end{picture}
     \vskip -0.2cm
    \caption{(a) The average circular polarization $\overline{\xi}_2$ (blue solid line) and average OAM (green dashed line) versus photon energy $\omega$ (MeV) for $\theta_\perp=0.72$ mrad. The OAM distributions of incoherent vortex $\gamma$ photons for the first (b) and second (c)  harmonics. The parameters are the same as in  Fig. \ref{Fig. agl}.}
    \label{Fig. alternative}
\end{figure}

{\color{black}{In addition to the distinct spectral helicity observed in Fig. \ref{Fig. agl} (c), the spectral OAM also reveals a pronounced structure in a two-color counter-rotating CP field; see Fig. \ref{Fig. m}.}}
%Now we proceed to explore the OAM property of the $\gamma$ photons in a two-color counter-rotating CP field. %The OAM number of the twisted photon can be calculated with $l=m-s$ govern by angular momentum conservation.
Energy conservation in the scattering process requires the 
$n$th harmonic photon energy to be $\omega\equiv n\omega_1=n_1\omega_1+n_2\omega_2$, where $n=n_1+2n_2$ with $n_1$ and $n_2$ being the numbers of photons absorbed from the RCP and LCP pulse, respectively, the fundamental frequency of emissions $\omega_{1}\backsimeq\frac{4\gamma_{0}^{2}\omega_{01}}{1+\gamma_{0}^{2}\theta_\perp^{2}+\left(a_{01}^{2}+a_{02}^{2}\right)/2}\approx 7$ MeV and $\omega_2=2\omega_1$ \cite{taira2018gamma}.  For the $n$th harmonic, there are multiple combinations of $n_1$ and $n_2$, corresponding to different photon absorption channels.  
%In the limit $\theta_\perp\ll1$, the Lorentz boosted frequency are related to $\omega_{01}$ and $\omega_{02}$ via \cite{taira2018gamma}
%\begin{align}
%\omega_{1}&\backsimeq\frac{4\gamma_{0}^{2}\omega_{01}}{1+\gamma_{0}^{2}\theta_\perp^{2}+\left(a_{01}^{2}+a_{02}^{2}\right)/2},\quad \omega_2=2\omega_1.
%\end{align}
%For $\gamma=2\times 10^3$, $\omega_{01}=1.55$eV, $\theta_\perp=0.72$mrad, $a_{01}=a_{02}=0.5$, we have $\omega_{1}\approx$ 7 MeV, which is coincide with the energy of the first peak in Fig \ref{Fig. agl} (c). 
Meanwhile, conservation of photon angular momentum requires the  transferred angular momentum satisfies $m=n_1-n_2$, which can be either positive and negative depending on the number of photons absorbed from each driver $(n_1,n_2)$. Consequently, each harmonic exhibits multiple OAM (SAM) distributions with varying signs; see Fig. \ref{Fig. m}.
Specifically, for the first harmonic with energy of $\omega\approx 7$ MeV, the electron absorbs one photon form RCP laser and emits a vortex photon with total angular momentum $m=1$. This vortex photon can be in one of two final states: $s=1,l=0$ and $s=-1,l=2$ (Fig. \ref{Fig. m}). Apparently, the emission of LCP photons with $l=2$ is most probable. 
For the second harmonic, there are two possible channels for producing photons with energy $\omega=14$ MeV. The electron can absorb either two RCP laser photons with $\omega_1$ or one LCP photon with $\omega_2$. The absorbed angular momentum for these channels is  $m=+2$ and $m=-1$, respectively. The first channel contributes to photon emissions with
 $s=1,l=1$  and $s=-1,l=3$, while the second channel contributes with $s=1,l=-2$ and $s=-1,l=0$. As shown in Fig. \ref{Fig. m}, the second harmonic is dominated by RCP photons with $l=-2$.
Similarly, the third harmonic has two possible channels: $\omega=\omega_1+\omega_2$ or  $\omega=3\omega_1$. The angular momentum transferred from the laser to the $\gamma$ photons is $m=0$ and  $m=3$, respectively. The first channel contributes to photon emissions with $s=1,l=-1$ and $s=-1,l=1$, while the second channel contributes with $s=1,l=2$ and $s=-1,l=4$. Among all the final states, the probability of emitting a LCP photon with $l=1$ is the highest.
For the fourth harmonic, there are three possible channels: $\omega=4\omega_1$, $\omega=2\omega_1+\omega_2$, and $\omega=2\omega_2$. In this case, multiple OAM modes can be found, with $l$ varying from -3 to 5 [Fig. \ref{Fig. m}]. The dominant component is the RCP photon with $l=-3$.
With the increase in harmonic order, the leading components vary as follows: $s=-1,l=2$; $s=+1,l=-2$; $s=-1,l=1$; and $s=+1,l=-3$. %Consequently, one can anticipate oscillations in photon polarization and vortex charge within the spectrum.

To characterize the vortex property of photons with multiple OAM modes, we define the average OAM as $\overline{l}(\omega)=\frac{\sum_i l_iP_i(\omega)}{\sum_i P_i(\omega)}$, where $P_i(\omega)$ represents the probability of emitting a photon with energy $\omega$ and topological charge $l_i$. Similarly, the average circular polarization can be defined as $\overline{\xi}_2(\omega) = \frac{P_+(\omega)-P_-(\omega)}{P_+(\omega)+P_-(\omega)}$, with $P_+(\omega)$ and $P_-(\omega)$ being the probability for emitting a RCP and LCP photons with energy of $\omega$, respectively. Oscillations in $\overline{\xi}_2$ and $\overline{l}$ are clearly evident in the spectrum [Fig. \ref{Fig. alternative} (a)].  
As the emitted photon energy increases, the polarization of the emitted photons alternates between RCP and LCP, while the topological charge shifts between negative and positive values.   %Additionally, this unique structure enables the generation of spatially isolated vortex beams of opposite spin and orbital angular momentum through a crystal diffraction system or multilayer mirrors, providing a way to obtain vortex photons with specific polarization states and OAM modes. 
The circular polarization can reach  $|\overline{\xi}_2|\sim 70\%$, and the topological charge can attain $|\overline{l}|\sim1.7$. {\color{black}{Therefore, we have successfully obtained vortex $\gamma$ photons with unique spectral structure,}} enabling the generation of highly polarized vortex photons with specific {\color{black}{helicity}} and twist directions through post-energy-selection techniques. Note that the $\gamma$-ray emissions consist of incoherent photons with various OAM modes. The OAM purity decreases with increasing harmonic order, as shown in Fig. \ref{Fig. alternative} (b) and (c). The first harmonic contains only two OAM modes, while the second harmonic features four OAM modes.

\begin{figure}[t]
   \includegraphics[width=0.5\textwidth]{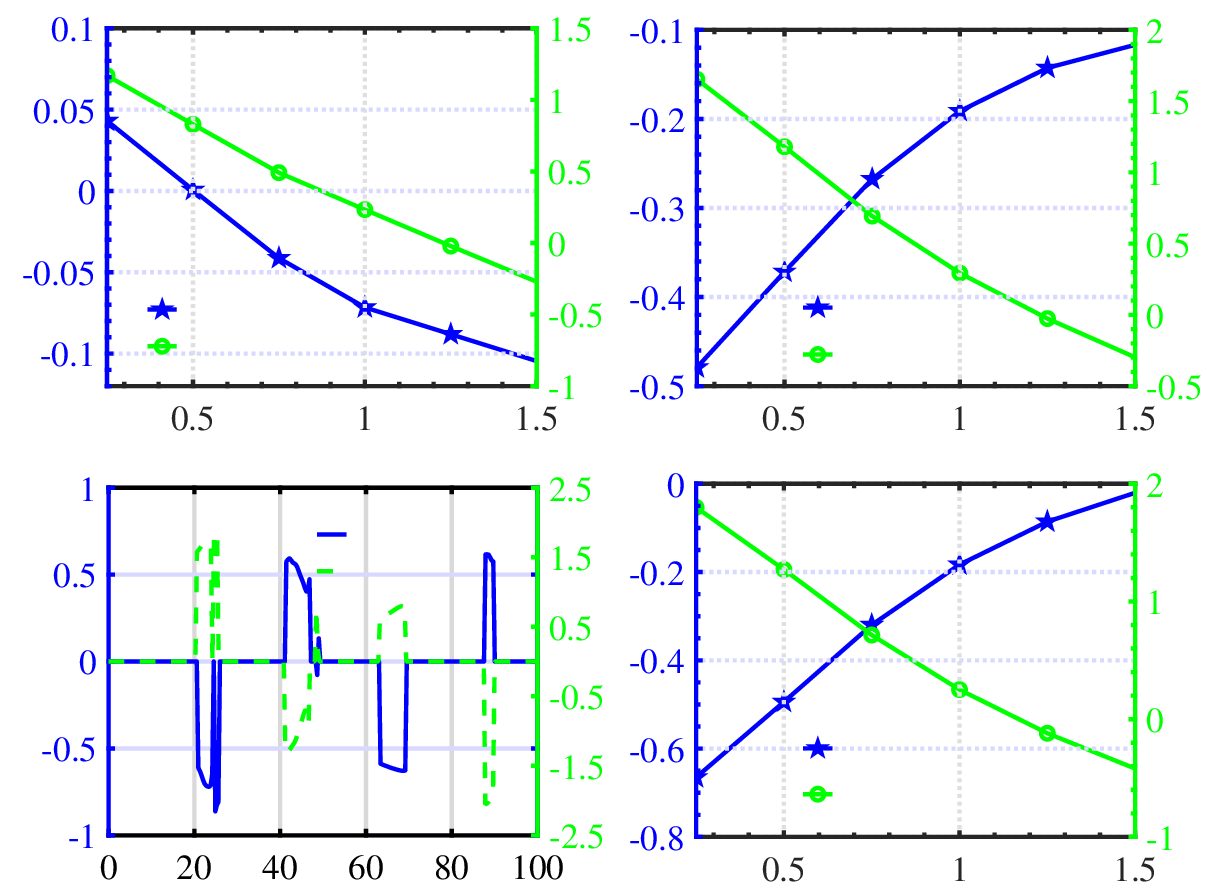}
     \begin{picture}(300,15)       
     \put(99,186){(a)}
     \put(41,135){\fontsize{7pt}{\baselineskip}\selectfont{$\overline{\xi}_2$}}
     \put(41,124){\fontsize{7pt}{\baselineskip}\selectfont{$\overline{l}$}}
     \put(55,104){\fontsize{9.0pt}{\baselineskip}\selectfont{$a_2/a_1$}}
     \put(99,91){(c)}
     \put(76,88){\fontsize{7pt}{\baselineskip}\selectfont{$\overline{\xi}_2$}}
     \put(76,78){\fontsize{7pt}{\baselineskip}\selectfont{$\overline{l}$}}
     \put(55,6){\fontsize{9.0pt}{\baselineskip}\selectfont{$\omega$(Mev)}}
     \put(148,186){(b)}
     \put(175,135){\fontsize{7pt}{\baselineskip}\selectfont{$\overline{\xi}_2$}}
     \put(175,124){\fontsize{7pt}{\baselineskip}\selectfont{$\overline{l}$}}
     \put(181,104){\fontsize{9.0pt}{\baselineskip}\selectfont{$a_2/a_1$}}
     \put(149,91){(d)}
     \put(175,43){\fontsize{7pt}{\baselineskip}\selectfont{$\overline{\xi}_2$}}
     \put(175,32){\fontsize{7pt}{\baselineskip}\selectfont{$\overline{l}$}}
     \put(181,6){\fontsize{9.0pt}{\baselineskip}\selectfont{$a_2/a_1$}}
      \end{picture}
      \vskip -0.2cm
    \caption{The scaling laws of the average  polarization $\overline{\xi}_2$ (blue line with star makers) and OAM $\overline{l}$ (green line with circle markers) versus relative intensity ratio $a_2/a_1$ for: (a) full angle, and (b) $\theta_\perp=0.72$ mrad. (c) The average circular polarization (blue solid line) and average OAM (green dashed line) versus photon energy $\omega$ (MeV) for $a_{01}=a_{02}=0.5$, $\varepsilon_0=2$GeV. (d) Same as (b) but for $\varepsilon_0=2$GeV and $\theta_\perp=0.43$ mrad. The intensity ratio is controlled by fixing $a_1=0.5$ and increasing $a_2$ from 0.125 to 0.75. Other parameters are same with Fig. \ref{Fig. agl}. }
    \label{Fig. scaling}
\end{figure}

{\color{black}{More importantly, our scheme}} facilitates easy tuning of the OAM of $\gamma$ photons by adjusting the intensity ratio of the two fields; see Fig. \ref{Fig. scaling}. As the relative ratio $R=a_2/a_1$ increases from $R=0.25$ to $R=1.5$, the average OAM of the emitted photons decreases from $\overline{l}=1.17$ to $\overline{l}=-0.27$. Simultaneously, the circular polarization of the emitted photons decreases from $\overline{\xi}_2=4.3\%$ to $-10.5\%$ [Fig. \ref{Fig. scaling} (a)]. As $R$ rises, the dominant photon absorption channel shifts from RCP to LCP, leading to a decrease in both $\overline{\xi}_2$ and $\overline{l}$. Meanwhile, the scaling law for polarization is dependent on the selection angle [Fig. \ref{Fig. scaling} (b)]. %For small $R$, the radiation at $\theta_\perp=0.72$ mrad is primarily dominated by the $s=-1$ components; see Figs. \ref{Fig. benchmark} (c) and (d) in Appendix.
As $R$ increases, the LCP laser contributes more $s=+1$ components to the large-angle region [Fig. \ref{Fig. agl} (a)], leading to the increase of polarization from $\overline{\xi}_2=-48\%$ to -12\% at $\theta_\perp=0.72$ mrad.
%For $\theta_\perp=0.72$ mrad, the emitted photons have a negative $\overline{\xi}_2$. 
In contrast, the scaling law of topological charge is robust against the variation of selection angle. As the relative ratio $R=a_2/a_1$ increases from $R=0.25$ to $R=1.5$, the average OAM of the emitted photons decreases from $\overline{l}=1.65$ to $\overline{l}=-0.3$. 
The average OAM can be estimated by $\overline{l}=n_1-n_2-\overline{\xi}_2\propto a_{01}^3-a_{02}^3$ for small $\overline{\xi}_2$ \cite{seipt2017depletion}.
When $a_2<a_1$, the average topological charge is positive due to the dominance of photon absorption from RCP laser photons. Conversely, when $a_2>a_1$, the average topological charge become negative as LCP laser photons dominate the absorption process.
Therefore, %the counter-rotating CP fields can generate $\gamma$ photons with continuously tunable polarization and topological charge.
by manipulating the laser intensity ratio, RCP or LCP photons with phase twisting in the clockwise or counterclockwise direction can be achieved.

For experimental feasibility, we investigate the impact of the initial electron energy on the controlling of SAM and OAM; see Figs. \ref{Fig. scaling} (c) and (d) {\color{black}{(see more details in Appendix D)}}. For an incident electron with an energy of 2 GeV, the spectrum retains characteristic oscillations in polarization and vortex charge observed in Fig. \ref{Fig. alternative} (a), but with a smaller opening angle $\theta_\perp=0.43$ mrad. The decreases of $\theta_\perp$ results from the enhanced collimation of $\gamma$-ray emissions at higher $\varepsilon_0$.   The circular polarization reaches  $|\overline{\xi}_2|\sim 86\%$, and the topological charge attains $|\overline{l}|\sim 2.1$.   Additionally, as the relative ratio $R=a_2/a_1$ increases from $R=0.25$ to $R=1.5$, the average OAM of the emitted photons decreases from $\overline{l}=1.8$ to $\overline{l}=-0.42$.  Simultaneously, the circular polarization of the emitted photons increases from $\overline{\xi}_2=-67\%$ to $-2\%$. The scaling law aligns with the trends observed in Fig. \ref{Fig. scaling} (b), demonstrating the robustness of our method to variations in initial electron energy.

In conclusion, we present a scheme of generating $\gamma$ beams with tunable SAM and OAM through NCS at moderate laser intensities. By employing a two-color counter-rotating CP fields, %we are able to manipulate the angular momentum exchange during the upconversion process, mapping the desired angular momentum of near-infrared laser light to $\gamma$-ray radiation. Our scheme 
we generates spectrally isolated vortex photons with opposing SAM and OAM, and provides a novel approach to control {\color{black}{helicity}} and twist directions through fine-tuning of the laser field intensities, which can be experimentally implemented by defocusing one of the laser pulses. Our study propelled the topic of optical angular momentum control from attosecond science to the forefront of strong-field QED. The obtained $\gamma$ vortex with controllable SAM and OAM may have potential applications in generating positron beam with controllable angular momentum, laboratory astrophysics, quantum information processing, high-precision nuclear manipulation and microscopy.

%For a given opening angle, different channels for twisted photon emission are analysed with energy and angular momentum conservations. %The resulting spectrum exhibits distinct oscillations in both spin and orbital angular momentum.  The polarization and topological charge of the twisted photons alternates between positive and negative values as the harmonic order increases. When such photons interact with atoms or nuclei, those at different depths may undergo varying rotational effects.  Additionally, this unique structure enables the generation of spatially isolated vortex beams of opposite spin and orbital angular momentum through a crystal diffraction system or multilayer mirrors, providing a way to obtain twisted photons with specific polarization states and OAM modes. Moreover, the average OAM of multiple modes and the polarization of the photons can be adjusted by tuning the relative intensities of the laser fields, which can be achieved simply by defocusing one of the laser pulses. With nonlinear Compton scattering of a two-color co-rotating  CP fields, we extended the control of polarization and vortex charge in the XUV to $\gamma$-ray regime, which may have applications in generating  positron beam with controllable angular momentum, laboratory astrophysics, quantum information processing, high-precision nuclear manipulation and  microscopy.

{\it Acknowledgement:}
This work is supported by the National Natural Science Foundation of China (Grants No.12474312, No. 12074262,  No.12425510, No. 12441506, and No.U2267204), the National Key R\&D Program of China (Grant No. 2021YFA1601700 and Grant No. 2024YFA1610900).

\vspace{10pt}

\appendix

\begin{figure}[]
    \includegraphics[width=0.465\textwidth]{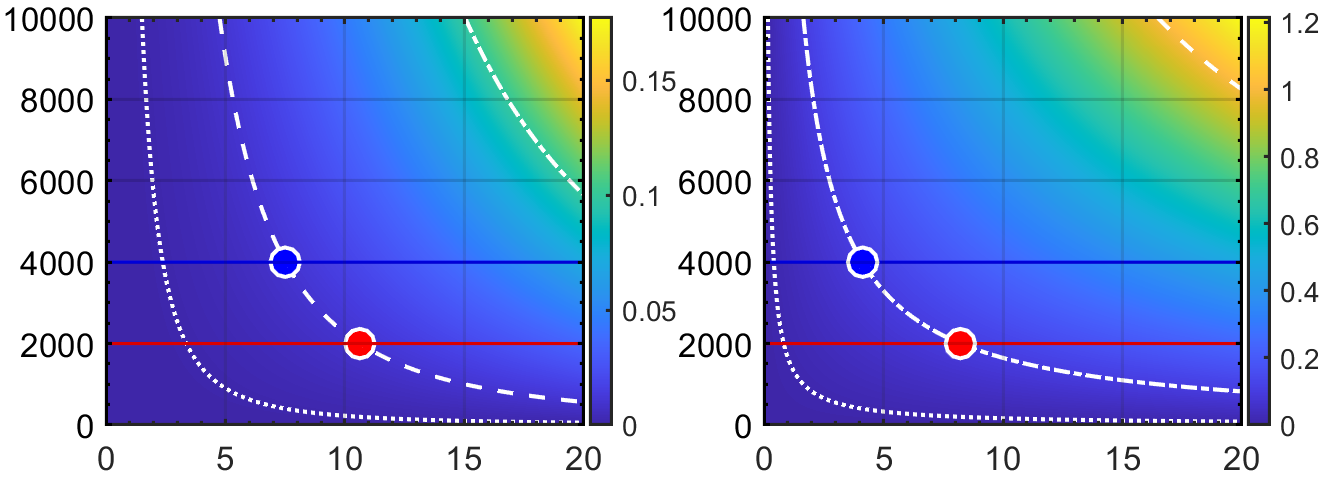}\\
     \begin{picture}(300,20)
    \put(33,97){\color{white}{\fontsize{8pt}{\baselineskip}\selectfont (a)}}
    \put(150,98){\color{white}{\fontsize{8pt}{\baselineskip}\selectfont (b)}}
     \put(74,62){\textcolor{white}{\rotatebox{0}{\fontsize{6.0pt}{\baselineskip}\selectfont 2Gev}}}
     \put(74,48){\textcolor{white}{\rotatebox{0}{\fontsize{6.0pt}{\baselineskip}\selectfont 1Gev}}}
     \put(25,36){\textcolor{white}{\rotatebox{0}{\fontsize{6.0pt}{\baselineskip}\selectfont Weak damping}}}
     \put(33,77){\textcolor{white}{\rotatebox{284}{\fontsize{6.0pt}{\baselineskip}\selectfont $R_c$=0.001}}}
     \put(47,89){\textcolor{white}{\rotatebox{291}{\fontsize{6.0pt}{\baselineskip}\selectfont $R_c$=0.01}}}
     \put(88,89){\textcolor{white}{\rotatebox{301}{\fontsize{6.0pt}{\baselineskip}\selectfont $R_c$=0.1}}}
     \put(52,99){\textcolor{white}{\rotatebox{0}{\fontsize{6.0pt}{\baselineskip}\selectfont Strong damping}}}
     \put(182,62){\textcolor{white}{\rotatebox{0}{\fontsize{6.0pt}{\baselineskip}\selectfont 2Gev}}}
     \put(182,48){\textcolor{white}{\rotatebox{0}{\fontsize{6.0pt}{\baselineskip}\selectfont 1Gev}}}
     \put(153,36){\textcolor{white}{\rotatebox{0}{\fontsize{6.0pt}{\baselineskip}\selectfont Classical}}}
     \put(143,76){\textcolor{white}{\rotatebox{276}{\fontsize{6.0pt}{\baselineskip}\selectfont $\chi_e$=0.01}}}
     \put(152,90){\textcolor{white}{\rotatebox{286}{\fontsize{6.0pt}{\baselineskip}\selectfont $\chi_e$=0.1}}}
     \put(210,98){\textcolor{white}{\rotatebox{314}{\fontsize{6.0pt}{\baselineskip}\selectfont $\chi_e$=1}}}
     \put(177,99){\textcolor{white}{\rotatebox{0}{\fontsize{6.0pt}{\baselineskip}\selectfont Quantum}}}
    \put(62,15){\fontsize{8pt}{\baselineskip}\selectfont$a_0$}
     \put(-2,69){{\rotatebox{90}{\fontsize{8pt}{\baselineskip}\selectfont $\gamma_0$}}}   
    \put(182,15){\fontsize{8pt}{\baselineskip}\selectfont$a_0$}
\end{picture}
\vskip -0.6cm
    \caption{(a) The parameter $R_c$ versus $a_0$ and $\gamma_0$ for an 800 nm wavelength laser. Classical radiation damping becomes strong when $R_c\gtrsim0.01$  and dominates when $R_c\gtrsim0.1$. The threshold at which the radiation reaction begins to be significant: $a_{0c}\approx 10$ for 1GeV electron (red dot) and $a_{0c}\approx 7$ for 2GeV electron (blue dot). (b) The parameter $\chi_e$ versus $a_0$ and $\gamma_0$ for an 800 nm wavelength laser.  Quantum corrections to the spectrum become necessary when $\chi_e\gtrsim0.1$. The threshold at which the quantum correction begins to be significant: $a_{0c}\approx 8$ for 1GeV electron (red dot)  and $a_{0c}\approx 4$ for 2GeV electron (blue dot)}
    \label{Fig. boundary}
\end{figure}

{\color{black}{
\section{Appendix A: Boundary of laser intensity for neglecting radiation reaction}
 The parameter \(\chi_e = 2\omega_0 a_0\gamma_0/m\) is the quantum parameter that allows us to characterize the importance of quantum effects, such as photon recoil and spin effects \cite{di2012extremely}. Generally, for \(\chi_e \lesssim 0.1\), or equivalently \(a_0 \lesssim 1.7 \times 10^4/\gamma_0\) (\(I \lesssim 6 \times 10^{26}/\gamma_0^2 \, \text{W/cm}^2\)), radiation reaction is insignificant; see Fig. \ref{Fig. boundary} (b). More rigorously, in the classical regime (\(\chi_e \ll 1\)), the magnitude of radiation losses can approximately be determined by the invariant classical radiation reaction parameter \(R_c = \alpha\chi_e a_0\) [Fig. \ref{Fig. boundary} (a)], which quantifies the energy radiated by the electron in a single cycle, i.e. \(R_c = 3\varepsilon_\text{rad}/4\pi\varepsilon_0\). If we define ``significant" radiation damping as an energy loss of approximately 10\% per period, the threshold is found to be \(R_c \approx 0.024\) \cite{thomas2012strong,blackburn2020radiation}. This corresponds to \(a_{0c} \approx 7 \times 10^2\gamma_0^{-1/2}\) (\(I_{0c} \approx 1 \times 10^{24}\gamma_0^{-1} \, \text{W/cm}^2\)) for a laser with a wavelength of \(800 \) nm. At this point, the force on the electron due to radiative losses should be included in the equations of motion. For GeV electrons, the threshold for strong damping is approximately \(a_{0c} \sim 10\) (\(I_{0c} \sim 10^{20} \, \text{W/cm}^2\)) [Fig. \ref{Fig. boundary} (a)], which is significantly larger than the \(a_0 = 0.5\) (\(I \sim 10^{17} \, \text{W/cm}^2\)) used in our setup. Therefore, radiation reaction effects are negligible in this context.  }}

{\color{black}{ 
\section{Appendix B: Definition of SAM and OAM}
 There is an on-going discussion on the definitions of SAM and OAM, stemming from the puzzling fact that the genuine gauge-invariant photon spin operator does not exist \cite{leader2014angular,palmerduca2024four,yang2022quantum}.
However, in our simulations, the SAM refers to the photon helicity, which is a gauge-invariant and well-defined operator \cite{weinberg1995quantum} and can be directly measured in experiments \cite{forbes2021measures}. Thus, we focus on the helicity as a physically meaningful and measurable quantity, rather than engaging directly with the broader debate surrounding OAM and SAM definitions.

For moderate laser fields (\( a_0 \sim 1 \)), the propagation axis of the scattered photons aligns with the reaction axis of the collision. In this case, the OAM quantization axis is uniquely defined \cite{ivanov2011scattering}. %and orbital helicity is approximately conserved \cite{ivanov2011scattering}. 
The detector is placed along the OAM quantization axis $\mathbf{e}_3$, which coincides with both the reaction axis of the collision and the average propagation direction of the emitted photons. 
In this work, we choose  $\mathbf{e}_3=(0,0,1)$ along the collision axis $z$, with the orthonormal vectors $\mathbf{e}_1=(1,0,0)$ and $\mathbf{e}_2=(0,1,0)$, forming a right-handed coordinate system. 
The quantum number $s$ defines the helicity of the vortex photon, $m$ the projection of the total angular momentum onto $\mathbf{e}_3$, $k_3$ and $k_\perp$ the projection of the photon momentum $k$ onto and orthogonal to $\mathbf{e}_3$, respectively. We define $n_3=k_3/k$ and $n_\perp=k_\perp/k$. The coordinates are represented as $\bm{x}_\pm(t)=(\mathbf{e}_\pm,\bm{x})$  and the velocities as $\bm{\upsilon}_\pm(t)=(\mathbf{e}_\pm,\bm{\upsilon})$, where $\mathbf{e}_\pm=\mathbf{e}_1\pm i\mathbf{e}_2$. The function $j_m$ is related to the Bessel function of the first kind $J_m$ through the relation $j_{m}\left(p,q\right)=\frac{p^{m/2}}{q^{m/2}}J_{m}\left(\sqrt{p},\sqrt{q}\right)$.

The radiation probability used in our simulation is derived using Bessel vortex states, which are the eigenstates of the total angular momentum operator \( \hat{J}_3 \) and the helicity operator \( \hat{S} \). This indicates that both helicity and total angular momentum along the \(z\)-axis can be simultaneously measured. Furthermore, for an incoming ultrarelativistic electron, the cone angle associated with the radiation is quite small, i.e., \( \theta \sim a_0/\gamma_0 \). This small-angle approximation (paraxial approximation) allows for interpreting the radiated photon's wave function as an eigenstate of both SAM and OAM operators with respective eigenvalues \( s \) and \( l = m - s \) \cite{ababekri2024vortex,ivanov2022promises}.
}}

\section{Appendix C:  Benchmark of our simulation code} 
\begin{figure}
 \includegraphics[width=0.5\textwidth]{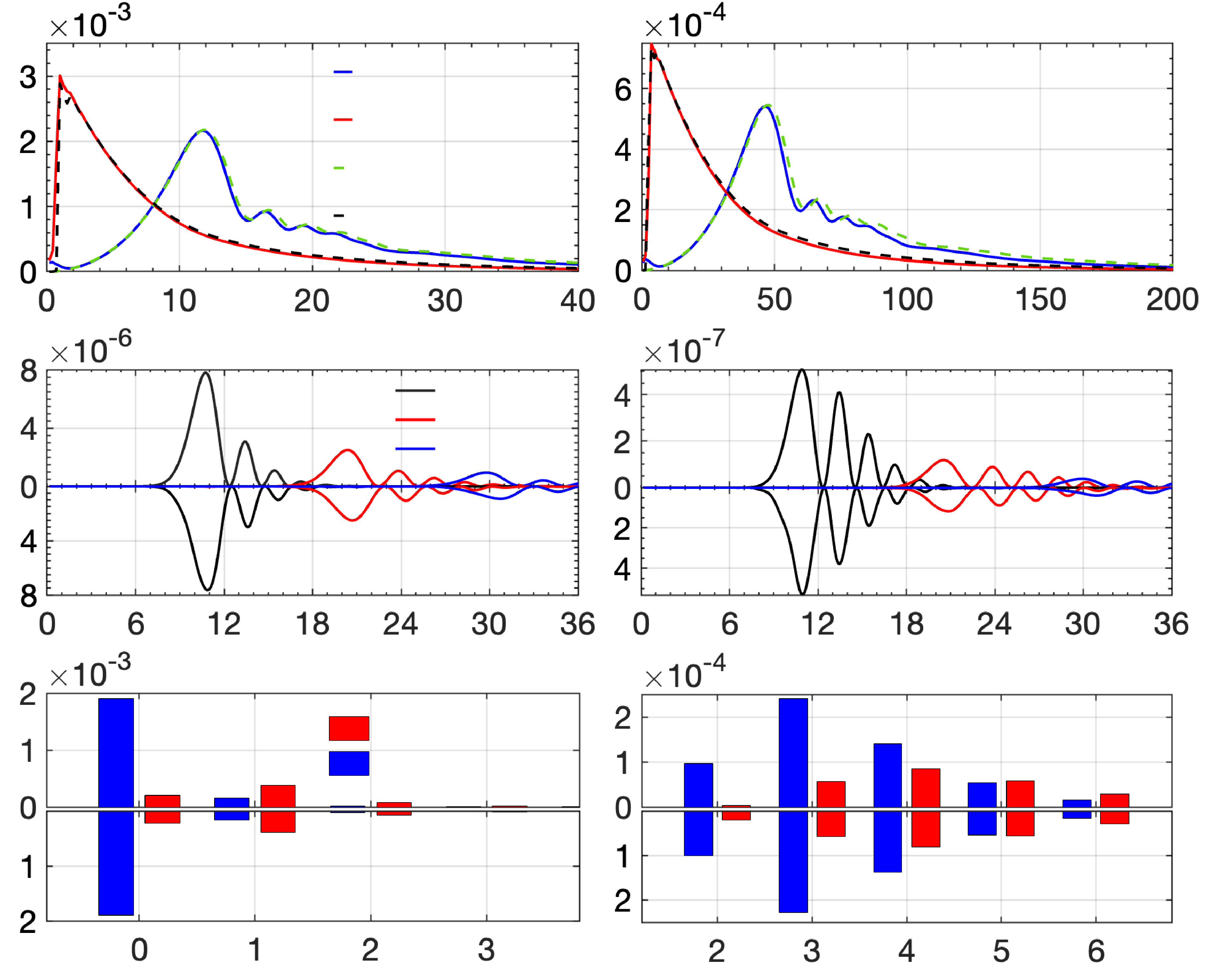}\\
     \begin{picture}(300,20)
    \put(112,213){\fontsize{7pt}{\baselineskip}\selectfont(a)}
     \put(238,213){\fontsize{7pt}{\baselineskip}\selectfont(b)}
     \put(112,144){\fontsize{7pt}{\baselineskip}\selectfont(c)}
     \put(238,144){\fontsize{7pt}{\baselineskip}\selectfont(d)}
     \put(112,76){\fontsize{7pt}{\baselineskip}\selectfont(e)}
     \put(239,75){\fontsize{7pt}{\baselineskip}\selectfont(f)}
     \put(56,155){\fontsize{7pt}{\baselineskip}\selectfont $\omega$(MeV)}  
     \put(180,155){\fontsize{7pt}{\baselineskip}\selectfont $\omega$(MeV)}
     \put(76,211){\textcolor{black}{\rotatebox{0}{\fontsize{6.0pt}{\baselineskip}\selectfont QED, $s$=+1}}}
     \put(76,201){\textcolor{black}{\rotatebox{0}{\fontsize{6.0pt}{\baselineskip}\selectfont QED, $s$=-1}}}
     \put(76,192){\textcolor{black}{\rotatebox{0}{\fontsize{6.0pt}{\baselineskip}\selectfont classical, $s$=+1}}}
     \put(76,182){\textcolor{black}{\rotatebox{0}{\fontsize{6.0pt}{\baselineskip}\selectfont classical, $s$=-1}}}   
     \put(-6,185){{\rotatebox{90}{\fontsize{7pt}{\baselineskip}\selectfont dP/d$\omega$}}}   
     \put(95,144){\textcolor{black}{\rotatebox{0}{\fontsize{6.5pt}{\baselineskip}\selectfont  $n$=1}}}
     \put(95,138){\textcolor{black}{\rotatebox{0}{\fontsize{6.5pt}{\baselineskip}\selectfont $n$=2}}} 
     \put(95,132){\textcolor{black}{\rotatebox{0}{\fontsize{6.5pt}{\baselineskip}\selectfont $n$=3}}}  
     \put(56,86){\fontsize{7pt}{\baselineskip}\selectfont $\omega$(MeV)}  
     \put(180,86){\fontsize{7pt}{\baselineskip}\selectfont $\omega$(MeV)}
     \put(-6,113){{\rotatebox{90}{\fontsize{7pt}{\baselineskip}\selectfont d$^2$P/d$\omega$d$\theta$}}}
     \put(81,72){\textcolor{black}{\rotatebox{0}{\fontsize{6.5pt}{\baselineskip}\selectfont 19MeV}}}
     \put(81,64){\textcolor{black}{\rotatebox{0}{\fontsize{6.5pt}{\baselineskip}\selectfont 12MeV}}}   
    \put(67,15){\fontsize{8pt}{\baselineskip}\selectfont $l$}
    \put(-6,48){{\rotatebox{90}{\fontsize{8pt}{\baselineskip}\selectfont dP/d$\omega$}}}   
    \put(190,15){\fontsize{8pt}{\baselineskip}\selectfont $l$}
\end{picture}\\
\vskip -0.6cm
 \caption{(Upper row) The radiation probability $\text{d}P/\text{d}\omega$ $ (\text{eV}^{-1})$ for vortex \(\gamma\) photons in an RCP laser field with initial electron energies of 1 GeV (a) and 2 GeV (b).  
(Middle row) The angle-resolved radiation probability  $\text{d}^2P/\text{d}\omega \text{d}\theta$ $ (\text{eV}^{-1} \text{rad}^{-1})$ for vortex \(\gamma\) photons, with \(\theta_\perp = 0.35 \, \text{mrad}\) for the first three harmonics for helicities (c) \(s = +1\) and (d) \(s = -1\).  
(Bottom row) OAM distributions of incoherent vortex \(\gamma\) photons with energies \(\omega = 12 \, \text{MeV}\) and \(\omega = 19 \, \text{MeV}\) for helicities (e) \(s = +1\) and (f) \(s = -1\). The upper lines/bars in (c)-(f) correspond to QED calculations from Ref. \cite{ababekri2024vortex}, while the lower lines/bars represent semiclassical calculations from Eq. (\ref{probability}).  Electron energy $\varepsilon_0=1$ GeV, the laser intensity $a_0=1$, the central laser photon energy $\omega_0=1.55$ eV, and pulse duration $\tau\approx26.7$ fs.}
 \label{compare}
     
\end{figure}

\begin{figure}[]
\includegraphics[width=0.5\textwidth]{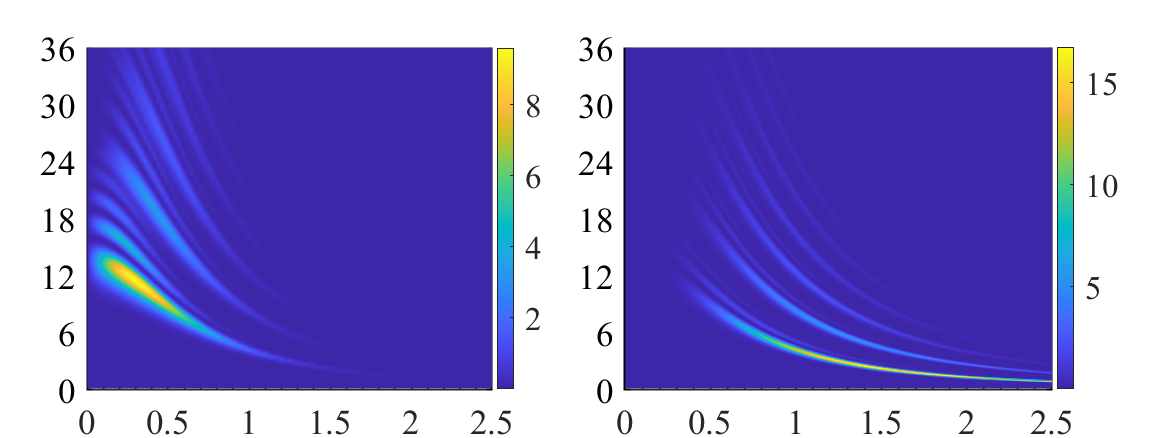}
\begin{picture}(300,20)       
     %\put(172,110){\fontsize{8.5pt}{\baselineskip}\selectfont$\theta_\perp$(mrad)}
     \put(96,98){\color{white}(a)} 
     \put(50,6){\fontsize{9pt}{\baselineskip}\selectfont$\theta_\perp$(mrad)} 
     \put(-4,49){\fontsize{8.5pt}{\baselineskip}\selectfont\rotatebox{90}{\fontsize{8.5pt}{\baselineskip}\selectfont $\omega$ (Mev)}}       
     \put(219,98){\color{white}(b)}
     \put(170,6){\fontsize{9pt}{\baselineskip}\selectfont$\theta_\perp$ (mrad)}
    \end{picture}
\vskip -0.2cm
 \caption{ The angle-resolved energy spectra $\omega\text{d}^2P/\text{d}\omega \text{d}\theta_\perp$  for photon helicities (a) $s=+1$ and (b) $s=-1$. The laser and electron parameters are same with Fig. \ref{compare}. }

    %\includegraphics[width=0.5\textwidth]{benchmark.eps}
    % \begin{picture}(300,15)       
    %  \put(100,196){(a)}
    %  \put(52,110){\fontsize{9pt}{\baselineskip}\selectfont$\omega$ (MeV)}
   %   \put(-5,155){\fontsize{9pt}{\baselineskip}\selectfont\rotatebox{90}{dP/d$\omega$}} 
   %   \put(68,192){\fontsize{7.5pt}{\baselineskip}\selectfont{$m=1$}} 
    %   \put(68,183){\fontsize{7.5pt}{\baselineskip}\selectfont{$m=2$}} 
     %  \put(68,174){\fontsize{7.5pt}{\baselineskip}\selectfont{$m=3$}} 
    %  \put(225,197){(b)}
     % \put(172,110){\fontsize{8.5pt}{\baselineskip}\selectfont$\omega$ (MeV)}
    %  \put(100,93){\color{white}(c)} 
     % \put(52,6){\fontsize{9pt}{\baselineskip}\selectfont$\theta_\perp$(mrad)} 
    %  \put(-8,49){\fontsize{8.5pt}{\baselineskip}\selectfont\rotatebox{90}{\fontsize{8.5pt}{\baselineskip}\selectfont $\omega$ (MeV)}}  
      
   %   \put(225,93){\color{white}(d)}
    %  \put(170,6){\fontsize{9pt}{\baselineskip}\selectfont$\theta_\perp$ (mrad)}
    %  \put(202,192){\fontsize{7.5pt}{\baselineskip}\selectfont{$m=1$}} 
    %   \put(202,183){\fontsize{7.5pt}{\baselineskip}\selectfont{$m=2$}} 
    %   \put(202,174){\fontsize{7.5pt}{\baselineskip}\selectfont{$m=3$}} 
   %  \end{picture}

  %  \caption{ Radiation probability for the vortex $\gamma$ photons in a RCP laser field, with $\theta_\perp=0.35$ mrad for the first three harmonics for helicities (a) $s=+1$ and (b) $s=-1$. The angle-resolved energy spectra $d^2P/d\omega d\theta_\perp$  for photon helicities (c) $s=+1$ and (d) $s=-1$. Electron energy $\varepsilon=1$ GeV, the laser intensity $a_0=1$, the central laser photon energy $\omega_0=1.55$ eV, and pulse duration $\tau\approx26.7$ fs.}
    \label{Fig. benchmark}
\end{figure}

The radiation probability of vortex photons can be calculated by being integrated either into a pre-existing code that computes the trajectories of charged particles or be used as a post-processing tool that computes Eq. (\ref{probability}) on a set of pre-calculated trajectories. In our simulation, we numerically solve the Lorentz equations and obtain the trajectory of the electron in two-color laser fields %[see Fig. \ref{Fig. scheme} (c)-(e)].
Then, we calculate the time-dependent  $I_{3}$ and $I_{\pm}$ for each  opening angle $\theta_\perp$, photon energy $\omega$, spin angular momentum $s$ and total angular momentum number $m$.
Finally, we substitute $I_{3}$ and $I_{\pm}$ into the $m$-resolved spectral probability given in Eq. (\ref{probability}) and accumulate over time to obtain the probability for vortex photon emission. The accuracy of our code has been benchmarked with the QED calculations in Ref. \cite{ababekri2024vortex}.
{\color{black}{We have plotted the spectral probabilities [Figs. \ref{compare} (a)-(d)], OAM distributions  [Figs. \ref{compare} (e)-(f)] and angle-resolved spectra [Fig. \ref{Fig. benchmark}] with the parameters used in Ref. \cite{ababekri2024vortex}. Our results are in good agreement with Figs. 2 and 3 in Ref. \citep{ababekri2024vortex}.
According to Refs. \cite{di2010quantum,ritus1985quantum,di2012extremely}, quantum radiation reaction could induce: (i) an increase of photon yield at low energies, (ii) a decrease of photon yield at high energies, and (iii) a shift of the maximum of the photon spectrum toward low photon energies. However, these differences are negligible for 1GeV electrons [Fig. \ref{compare} (a)] and rather small for 2 GeV electrons  [Fig. \ref{compare} (b)].  For a definite opening angle (e.g. $\theta$=0.35mrad), the deviations between classical and QED calculations is a slight nonlinear redshift in higher frequencies %and a slight modification in the amplitude  
 [Figs. \ref{compare} (c) and (d)], consistent with the conclusion from Ref.  
 \cite{seipt2011nonlinear}. 
%It is obvious that the redshift is much more pronounced at higher frequencies. 
It is important to note that the electron energy should not exceed \( \gamma_0 = 10^4 \) for the laser intensity considered, as beyond this point, the differences induced by quantum radiation reaction become significant, and the classical description no longer holds \cite{seipt2011nonlinear}. Meanwhile, we present the OAM spectra with defined energy and polarization, shown in Figs. \ref{compare} (e) and (f), which are in good agreement with Fig. 3 in Ref. \cite{ababekri2024vortex}.}}

In single-{\color{black}{helicity}} fields, photons with opposite helicity to the laser have a larger opening angle $\theta_\perp$ and lower photon energy $\omega$ than those with the same helicity to laser. Therefore, in an RCP field, the photon emissions with $s=-1$  [Fig. \ref{Fig. benchmark} (a)] have a larger $\theta_\perp$ and a smaller $\omega$ than those with the $s=1$ [Fig. \ref{Fig. benchmark} (b)] in an RCP field. %In the two-color counter-rotating CP fields, the introduction of second harmonic LCP laser photons causes $s=1$ photons to extend into the low-energy, large-angle region, while $s=-1$ photons shift into the high-energy, small-angle region. 

{\color{black}{
\section{Appendix D: Impact of electron and laser parameters} The scattering of ultrarelativistic electrons (46.6 GeV) off Terawatt-class lasers ($a_0=0.3$) has been conducted by pioneering experiment E-144 at SLAC in the 1990s \cite{bula1996observation,bamber1999studies}, with recent experimental proposals to repeat a similar experiment at FACET-II \cite{chen2022preparation,yakimenko2019facet} and DESY \cite{abramowicz2019letter,jacobs2022luxe} in the highly nonlinear nonperturbative regimes. Our proposals can be implemented with a comparable experimental setup, but with electrons of significantly lower energy.
Meanwhile, our scheme is robust with respect to variations in initial electron energy. The typical energy spread of electrons from laser wakefield acceleration (LWFA) is between 5\% and 15\%, with recent experiments achieving an energy spread of around 1\% \cite{mirzaie2024all}. The scaling laws for varying energy are largely consistent [see Fig. \ref{105GeV}], suggesting that conducting scattering experiments with LWFA-generated electrons could be a feasible approach. 
All-optical Thomson/Compton scattering has been conducted worldwide \cite{poder2018experimental,cole2018experimental,mirzaie2024all,
yan2017high,chen2024platform}. These experiments did not observe the harmonic structure because the spectra were measured in the strong-field regime. As the laser intensity increases, the distance between successive harmonic peaks decreases, ultimately leading to a continuous spectrum that no longer exhibits a distinct harmonic structure \cite{mackenroth2011nonlinear}. %Additionally, for \(a_0 > 10\), the formation length \(l_f = \lambda_0 / a_0\) becomes much smaller than the laser wavelength, resulting in incoherent gamma photon radiation. 
However, the harmonic structure could emerge for a less intense laser with \(a_0 \lesssim 1\), as shown in Figs. \ref{compare} (c)-(d).

\begin{figure}[]
 \includegraphics[width=0.5\textwidth]{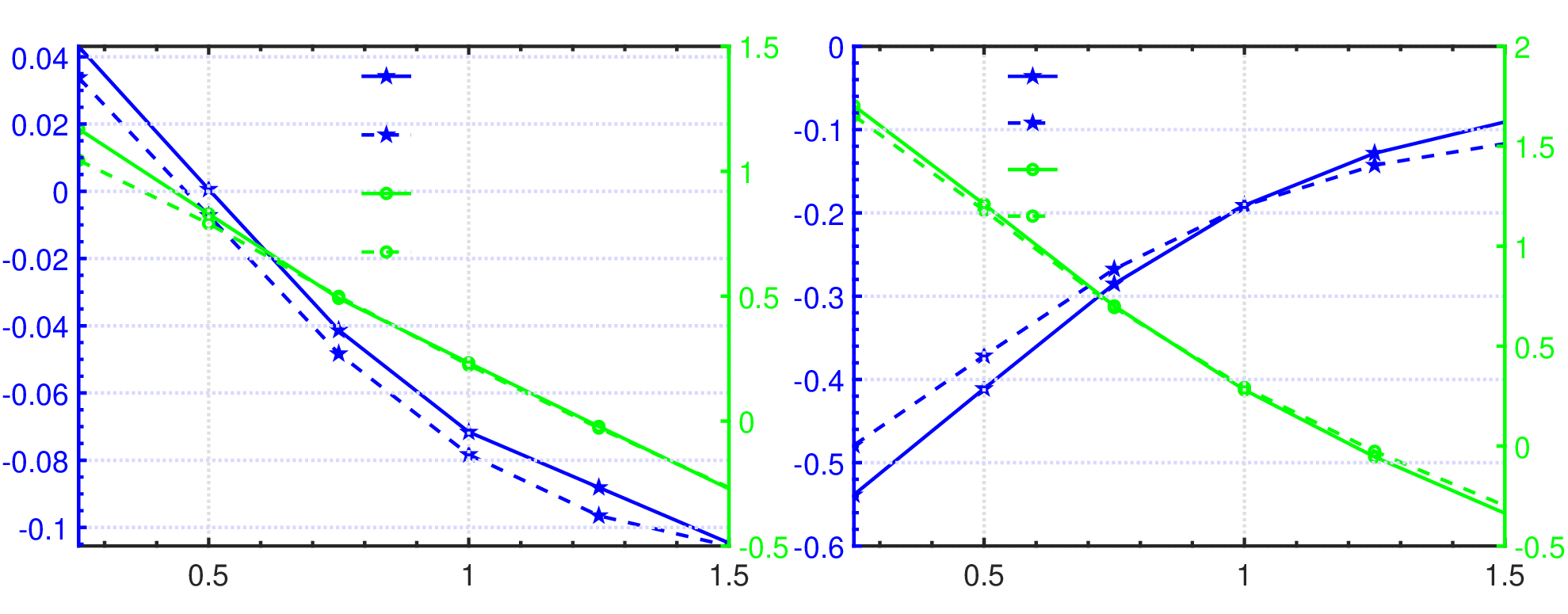}\\
     \begin{picture}(300,20)
    \put(107,106){\fontsize{7pt}{\baselineskip}\selectfont(a)}  
    \put(233,106){\fontsize{7pt}{\baselineskip}\selectfont(b)}  
    \put(69,107){\fontsize{6.0pt}{\baselineskip}\selectfont 1GeV, $\overline{\xi}_2$}
     \put(69,97){\fontsize{6.0pt}{\baselineskip}\selectfont 1.05GeV, $\overline{\xi}_2$}
     \put(69,88){\fontsize{6.0pt}{\baselineskip}\selectfont 1GeV, $\overline{l}$}
     \put(69,78){\fontsize{6.0pt}{\baselineskip}\selectfont 1.05GeV, $\overline{l}$}
     \put(170,107){\fontsize{6.0pt}{\baselineskip}\selectfont 1GeV, $\overline{\xi}_2$}
     \put(170,99){\fontsize{6.0pt}{\baselineskip}\selectfont 1.05GeV, $\overline{\xi}_2$}
     \put(170,91){\fontsize{6.0pt}{\baselineskip}\selectfont 1GeV, $\overline{l}$}
     \put(170,84){\fontsize{6.0pt}{\baselineskip}\selectfont 1.05GeV, $\overline{l}$}
    \put(55,19){\fontsize{7pt}{\baselineskip}\selectfont $a_2$/$a_1$}
    \put(180,19){\fontsize{7pt}{\baselineskip}\selectfont $a_2$/$a_1$}
\end{picture}\\
\vskip -0.7cm
\caption{The scaling laws of the average polarization $\overline{\xi}_2$ (blue line with star makers) and OAM $\overline{l}$ (green line with circle markers) versus
relative intensity ratio $a_2/a_1$ for: (a) full angle, and (b) $\theta_\perp$ = 0.72 mrad. }
\label{105GeV}
\end{figure}

We employed lasers with moderate intensity, \(a_0 = 0.5\) (\(I \approx 5 \times 10^{17} \, \text{W/cm}^2\)), which typically have a significantly larger beam size compared to the overlap mismatch caused by pointing-angle jitter of the lasers. For instance, to achieve an intensity of \(a_0 = 0.5\), a 1 PW laser (e.g., ELI-Beamlines, 30 J, 30 fs \cite{rus2017eli})  requires a beam size of \(w_0 \sim 10^2 \lambda\), while a 10 PW laser (e.g., the ELI-NP HPLS, 230 J, 23 fs \cite{tanaka2021status}) requires a beam size of \(w_0 \sim 10^3 \lambda\). Additionally, the jitter of high-power lasers is typically around $\sim$5 mrad (\(\sim \mu \text{m}\)) \cite{yan2017high}, which is much smaller than the laser beam spot size.  Therefore, the imperfect overlap of the two lasers due to pointing-angle jitter is  expected to be manageable for moderate laser intensities, making it feasible to tune the \(a_0\) ratios of the two lasers involved in the experiment. }}

\bibliography{prp}

\end{document}